\documentclass[twocolumn,prb,showpacs]{revtex4}

\usepackage{psfrag}
\usepackage{graphicx}
\usepackage{amsmath}
\usepackage{amssymb}
 
\begin{document}
\title{Image-charge induced localization of molecular orbitals at metal-molecule interfaces: Self-consistent GW calculations}
\author{M. Strange}
\affiliation{Center for Atomic-scale Materials Design, Department of
Physics \\ Technical University of Denmark, DK - 2800 Kgs. Lyngby, Denmark}
\author{K. S. Thygesen}
\email{thygesen@fysik.dtu.dk}
\affiliation{Center for Atomic-scale Materials Design, Department of
Physics \\ Technical University of Denmark, DK - 2800 Kgs. Lyngby, Denmark}
\date{\today}
 
\begin{abstract}
  Quasiparticle (QP) wave functions, also known as Dyson orbitals,
  extend the concept of single-particle states to interacting electron
  systems. Here we employ many-body perturbation theory in the GW
  approximation to calculate the QP wave functions for a
  semi-empirical model describing a $\pi$-conjugated molecular wire in
  contact with a metal surface. We find that image charge effects pull
  the frontier molecular orbitals toward the metal surface while
  orbitals with higher or lower energy are pushed away. This affects
  both the size of the energetic image charge shifts and the coupling
  of the individual orbitals to the metal substrate. Full
  diagonalization of the QP equation and, to some extent,
  self-consistency in the GW self-energy, is important to describe the
  effect which is not captured by standard density functional theory
  or Hartree-Fock. These results should be important for the understanding and theoretical modeling 
  of electron transport across metal-molecule interfaces.
 
\end{abstract}
\maketitle

\section{Introduction}
The independent-particle approximation and the associated one-electron orbital picture forms the basis of our understanding of chemical
bonding and electronic energy levels in solids and molecules. The most
widely used approximations of this type are Hartree-Fock (HF) and
density functional theory (DFT)\cite{HF,kohn-sham}. Although the single-particle orbitals derived from such schemes do not have physical meaning, apart from the fact that the exact DFT orbitals generate the exact groundstate density, they are routinely used to calculate and interpret physical quantities of various types. In strongly correlated systems such an approach clearly breaks down. However, even in weakly correlated systems where the single-particle picture is valid, there is no guarantee that the orbitals generated by the standard one-electron schemes are those which best resembles the true many-body excitations.

Quasiparticle (QP) wave functions provide a rigorous generalization of the concept of single-particle orbitals to interacting electron systems.
The QP states and energies are solutions of the QP equation\cite{hedin}
\begin{equation}\label{eq.QPeq}
[H_0 + \Sigma_{xc}(\varepsilon_\mu)]|\psi_{\mu}\rangle = \varepsilon_\mu|\psi_{\mu}\rangle
\end{equation}
Here $\hat H_0$ is the non-interacting part of the Hamiltonian including the
Hartree field while $\Sigma_{xc}$ is the non-local and
energy-dependent exchange-correlation (xc) self-energy operator\cite{comment}.
The QP energies represent the possible energies of a particle (electron or hole) added to the $N$-particle groundstate, and the QP wave function describes the probability amplitude for finding the added particle at a given position. (A precise definition and interpretation of the QP energies and wave functions will be given later in this paper.) 

The GW approximation\cite{hedin} to $\Sigma_{xc}$ (both in its self-consistent and non self-consistent form) has been succesfully used
to calculate QP energies of solids\cite{hybertsen,faleev,kresse,rinke,orr}, molecules\cite{GWmolecules,GWblase}, and more recently
solid-molecule interfaces\cite{neaton,juanma,prl_2009_thygesen,rinke2,biller}. The latter class of systems is particularly
challenging to describe due to its highly inhomogeneous nature:
Since the QP states describe the charged excitations, the QP
energies of an adsorbed molecule are strongly affected by the metal surface
through long range polarization effects (image charge effects) which decay as $1/z$ with $z$
being the distance to the surface, see Fig. 1(a). The inability of any available DFT functional to account for this renormalization of molecular energy levels, reflects the highly non-local nature of the phenomenon. We stress that this does not imply that the single-particle picture is invalid in such cases; only that the correct (QP) orbitals and energies cannot be obtained from a simple mean field potential.

\begin{figure}[h!]
  \includegraphics[width=0.9\linewidth]{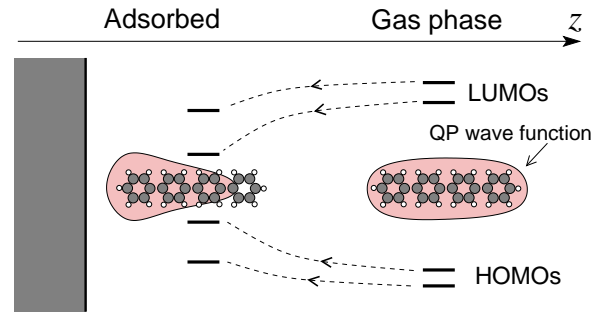}
  \caption{\label{fig0}(Color online) Schematic illustration of the change in the frontier QP energies and wave functions 
of a molecule approaching a metal surface. The closing of the HOMO-LUMO gap due to image charge screening is associated with a change in the shape of the orbital.}
\end{figure}
 
Most applications of the QP equation have focused on the QP
\emph{energies} while the QP wave functions have been much less studied. In fact it
is very often assumed that the latter, apart from normalization, are
identical to the orbitals obtained from DFT ($\psi_\mu^0)$. Under that
assumption, QP energies can be obtained from first order
pertubation theory involving only the diagonal matrix elements
$\langle \psi_\mu^0|\Sigma_{xc}(\varepsilon_\mu^0)|\psi_\mu^0\rangle$
thus greatly reducing the computational cost of solving the QP
equation. 

As an example where the shape of the QP wave functions play a key role
we consider the case of electron transport through a molecule
connected to metallic electrodes. The conductance of the junction
depends mainly on two factors, namely (i) the position of the
molecule's frontier (QP) energy levels relative to the metal Fermi
energy and (ii) the overlap between the molecule's frontier (QP)
orbitals and the extended states in the electrodes. The importance of
(i) has been studied in detail using scissors operator techniques to
correct the DFT energy levels while keeping the DFT orbitals
fixed\cite{quek,mowbray}. In contrast, the question of how well the
DFT orbitals resemble the true QP orbitals and the consequences for
charge transport has only been indirectly
studied\cite{strange,beilstein,rangel,neaton2,ferretti}.

In this paper we show, using many-body perturbation theory in the GW
approximation, that the QP wave functions of a molecular
wire in contact with a metal surface can be qualitatively different
from those obtained from an independent-particle approximation. While
the latter method yields orbitals that remain delocalized over the
molecule upon coupling to the surface, our GW calculations show that image charge effects not only
causes a reduction of the QP energies as previously
demonstrated, but also
renormalize the molecular orbitals. Orbitals with energy close to the
Fermi level are pulled more towards the surface while higher and lower
lying orbitals are pushed away. As a result, not only
the energies but also the life-times (inverse spectral width) of the
molecular resonances are affected by the xc self-energy.

The paper is organized as follows. In Sec. \ref{sec.QP} we briefly review the general quasiparticle theory including a transparent definition of the QP states and a simple proof of the equivalence between this definition and the QP equation. In Sec. \ref{sec.model} we introduce the metal-molecule interface model and in Sec. \ref{sec.method} we explain the method used to calculate the QP energies and wave functions. In Sec. \ref{sec.results} we present the results and discuss implications for modeling of charge transport in molecular junctions.

\section{Quasiparticle theory} \label{sec.QP} 
In this section we review the concept of the QP wave function and discuss its physical
meaning. For simplicity we shall make the assumption that the system under consideration is finite and the relevant excitations are discrete\cite{continuous}. 

We denote the $N$-particle groundstate and the excited states
by $|\Psi^N_0\rangle$ and $|\Psi^N_\mu\rangle$, respectively. The
occupied and unoccupied QP orbitals are denoted
$|\psi^{\text{-}}_{\mu}\rangle$ and $|\psi^{\text{+}}_{\nu}\rangle$,
respectively. The QP orbitals belong to the single-particle Hilbert
space and are defined through their matrix elements with a general
orbital $|\phi\rangle$:
\begin{eqnarray}\label{eq.QP1}
  \langle \phi|\psi^{\text{-}}_{\mu}\rangle^* &=&\langle \Psi_\mu^{N-1}|\hat c_\phi|\Psi_0^N \rangle\\ \label{eq.QP2}
\langle \phi|\psi^{\text{+}}_{\nu}\rangle &=&\langle \Psi_\nu^{N+1}|\hat c_\phi^{\dagger}|\Psi_0^N \rangle
\end{eqnarray}
where $\hat c_\phi$ and $\hat c_\phi^{\dagger}$ annihilates and creates an electron in the orbital $|\phi\rangle$, respectively. The real space representation of the QP wave functions are obtained by setting $|\phi\rangle =|r\rangle $ in the above equations. The QP wave functions defined above are also sometimes referred to as Lehman amplitudes or Dyson orbitals.

The QP energies are defined by 
\begin{eqnarray}\label{eq.QPen1}
\varepsilon^{-}_{\mu}&=& E^N_0-E^{N- 1}_\mu\\ \label{eq.QPen2}
\varepsilon^{+}_{\nu}&=& E^{N+ 1}_\nu-E^N_0
\end{eqnarray}
They represent the excitation energies of the $N\pm 1$ relative to $E^N_0$ and thus correspond to electron addition/removal energies. 

The definition of the QP wave functions given in Eqs. (\ref{eq.QP1},\ref{eq.QP2}) is not very transparent at first sight. 
A more transparent definition of the QP states can be obtained by noting that the projection 
\begin{equation}
\frac{|\langle \Psi_\mu^{N+1}|\hat c_\phi^\dagger|\Psi_0^N
\rangle|^2}{\langle \phi|\phi\rangle},
\end{equation}
is maximized exactly when $|\phi\rangle$ equals
$|\psi^{\text{+}}_{\mu}\rangle$.  In other words, $|\psi^{\text{+}}_{\mu}\rangle$ is the
orbital that makes $\hat c_\phi^\dagger|\Psi_0^N \rangle$ the
best approximation to the excited state $|\Psi_\mu^{N+1}\rangle$. Similarly, $|\psi^{\text{-}}_{\mu}\rangle$ is the orbital that makes $\hat c_\phi|\Psi_0^N \rangle$ the best approximation to the excited state $|\Psi_\mu^{N-1}\rangle$.
Consequently, the QP wave function is the single-particle orbital that best describes the state of the "extra" electron/hole in the excited state $|\Psi_\mu^{N\pm 1}\rangle$. In general, the QP states are non-orthogonal and their norm lies between 0 and 1. The norm is a measure of how well the excited many-body state can be described as a single-paticle excitation from the groundstate. 

In the special case of non-interacting electrons, the QP wave functions have norms exactly 1 or 0. The former correspond to excitations where one extra particle has been added to the groundstate Slater determinant. In this case the QP wave functions coincide with the normalized eigenstates of the one-electron Hamiltonian. The QP states with zero norm correspond to all other types of excitations.

It should be noted that the term "quasiparticle state" is often used only for those $|\psi^{\text{+/-}}_{\mu}\rangle$ whose norm is close to 1 while other states (those corresponding to collective excitations) are referred to as "satellites". In the present work we shall only consider QP states with norms very close to 1.

In the case where $E^{N\pm 1}_\mu$ belongs to the discrete spectrum of
the many-body Hamiltonian, it can be shown that
$\psi^{\pm}_{\mu}$ and $\varepsilon^{\pm}_{\mu}$ are solutions to the
QP equation (\ref{eq.QPeq}). In this case the norm of the QP state is
given by
$Z=(1-d\Sigma_{xc}(\varepsilon^{\pm}_{\mu})/d\varepsilon)^{-1}$. The
definition of QP states belonging to the continuum is a bit more
tricky\cite{continuous}, but this has no consequences for the present
work.

\section{Model}\label{sec.model}
\begin{figure}[h!]
  \includegraphics[width=0.9\linewidth]{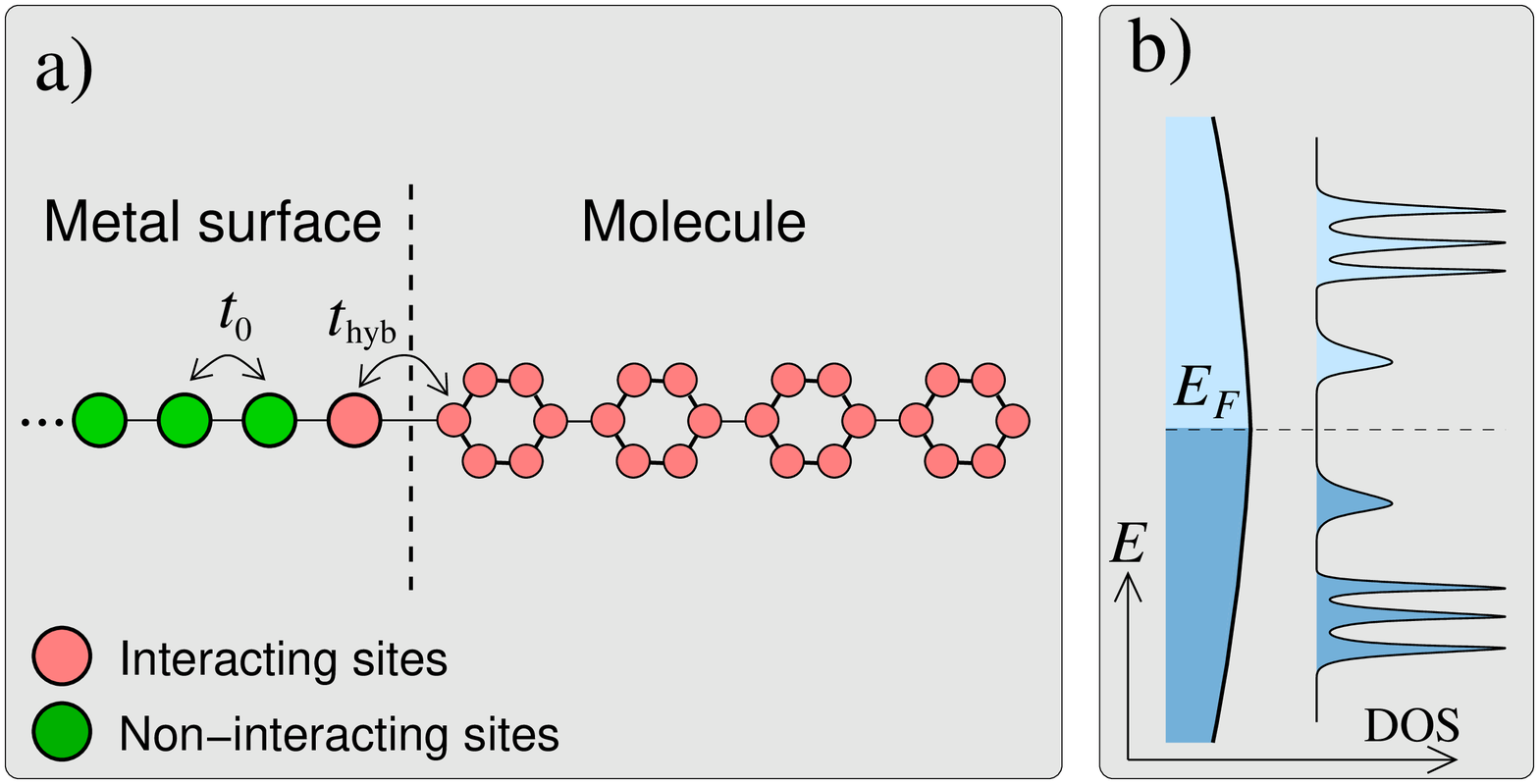}
  \caption{\label{fig1a}(Color online) a) Lattice model of a molecular wire interacting with a metal surface via hopping and Coulomb interaction.
b) The QP density of states (spectral function) of the adsorbed molecule. The different broadening of the levels reflects the difference in the shape of the QP orbitals.}
\end{figure}

We consider a four-unit paraphenylene molecular wire connected to a metal surface, see Fig. \ref{fig1a}. The paraphenylene molecule is described by a Pariser-Parr-Pople (PPP) Hamiltonian\cite{ppp_model}
\begin{equation}\label{eq.ham}
\hat{H}_\pi=\sum_{\langle ij \rangle, \sigma} t\hat{c}_{i\sigma}^\dagger \hat{c}_{j\sigma}+\frac{1}{2}\sum_{ij,\sigma\sigma'}V_{ij}\hat{c}_{i\sigma}^\dagger\hat{c}_{j\sigma'}^\dagger\hat{c}_{j\sigma'}\hat{c}_{i\sigma},
\end{equation}
where $\hat{c}^\dagger_{i\sigma}$($\hat{c}_{i\sigma}$) creates (annihilates) and electron at site $i$ ($p_z$ orbital of carbon atom $i$) with spin $\sigma$. The first term describes nearest neighbor hopping of strength $t=-2.4$ eV. In the second term, $V_{ij}$ is the long range Coulomb interaction acting between all sites on the molecule and for which we use Ohno's parametrization\cite{ohno}
\begin{equation}\label{eq.Vij}
V_{ij}=\frac{14.4}{\sqrt{(14.4/U)^2+ R_{ij}^2)}}, 
\end{equation}
where $R_{ij}$ is the distance between atoms $i$ and $j$ (in \AA) and $U=11.26$ (in eV). We note that the PPP model with the parameters used here in general provides an accurate description of the low-lying excitations in $\pi$ conjugated systems\cite{barford}. 
  
As a qualitative model of the metal surface we use a semi-infinite one
dimensional tight binding lattice. We use a large hopping parameter of
$t_0=-5.0$ eV between the sites of the chain to simulate a broad
featureless band and we set $E_F=0$ corresponding to half-filling. The last site on
the metallic chain is coupled to the nearest carbon atom of the
molecule by the hopping parameter $t_{hyb}=1.0$ eV (see inset of Fig.
\ref{fig2}). Coulomb interactions, $V_{ij}$, as defined in Eq.
(\ref{eq.Vij}) are included between the last site on the chain and all
the sites of the molecule. We set the distance between the last site
of the chain and the contacting carbon atom to 3 \AA. With these
parameters, the model yields realistic image charge shifts in the range 0.2-1.0 eV depending on the spatial form of the orbital\cite{quek,mowbray}. We stress that the Fermi energy lies in the
middle of the gap between the highest occupied molecular orbital (HOMO) and
lowest unoccupied molecular orbital (LUMO) such that the contacted
molecule remains in a closed-shell configuration far from the strong
correlation Kondo regime. 

We note that the use of a 1D chain to simulate the metal surface is clearly not adequate for quantitative computations. In particular, it cannot be used to describe the reorganization of electrons in the metal surface. However, from the viewpoint of the molecule it captures, at least qualitatively, all the aspects of the image charge effect in a real metal-molecule junction. In particular, the effect that a charge added to the molecule induces an image charge in the metal (change in the occupation of the last site of the 1D chain) which acts back on the electrons of the molecule. We also note that Delta-SCF as well as \emph{ab-initio} GW calculations for a benzene-diamine molecule adsorbed on an adatom, or small pyramid tip structure, on a gold surface, have shown that the image charge induced in the metal is largely confined to the adatom and is thus quite localized, see Figure 1 in Ref. \onlinecite{beilstein}.

\section{Method} \label{sec.method}
 
To obtain the QP wave functions and energies we calculate the
single-particle Green function following the method
described in Ref. \onlinecite{strange}. Briefly, the Green function of the contacted molecule is calculated from
\begin{equation}
G_{ij}(\omega)=[\omega-H_0-\Sigma_{\text{hyb}}(\omega)-\Sigma_{xc}(\omega)]^{-1}_{ij}.
\end{equation}
where $H_0$ is the non-interacting part of the molecular Hamiltonian including the Hartree field and $\Sigma_{\text{hyb}}$ is an embedding self-energy accounting for the coupling to the semi-infinite chain. In this work the xc self-energy is evaluated using either the Hartree-Fock or the GW approximation. Unless explicitly stated, the GW self-energy is evaluated fully self-consistently. The energy dependence of $G$ and $\Sigma_{\text{GW}}$ is sampled on a uniform grid $\omega_n=\varepsilon_n+i\eta$ where $\eta=0.01$ is an imaginary infinitesimal and $\varepsilon_n$ ranges from (in eV) -100 to 100 with a spacing of $\eta/2=0.005$.

We have previously shown that the GW approximation yields QP energies of molecules described by PPP models in good agreement with exact diagonalization results with an average deviation of the lowest QP energies of less than 5\%\cite{kaasbjerg}. In that work we also showed, using a measure for the degree of correlation based on the entropy of the reduced density matrix, that PPP models are significantly less correlated than Hubbard models with the same interaction strengths (obtained by removing all long-range interactions $V_{ij}$ with $i\neq j$ from the PPP model), explaining earlier studies which concluded that GW does not perform well for Hubbard clusters\cite{godby1,godby2}.

The Green function is related to the QP states and energies via its Lehmann representation\cite{ferdi}. Using this representation, the spectral function, $A(\omega)=(i/\pi)[G(\omega+i\eta)-G(\omega-i\eta)]$, projected onto the sites ($i,j$) of the molecule can be written as
\begin{equation}
A_{ij}(\omega)=\sum_{s\in\{+,-\}}\sum_\mu \langle i |\psi_\mu^{s}\rangle \langle \psi_\mu^{s}|j\rangle \delta(\omega-\varepsilon_\mu^s)
\end{equation}
We identify the molecular QP energies, $\varepsilon_n^{\text{mol}}$, of the molecule as the peaks in $A(\omega)$.
The corresponding QP orbital (precisely, the projection of the
QP orbital of the infinite metal-molecule system onto the molecule) is obtained as the unique solution to the eigenvalue equation
\begin{equation}
\sum_{j\in \text{mol}}A_{ij}(\varepsilon_n^{\text{mol}})\langle j|\psi_n^{\text{mol}}\rangle=\lambda
\langle i |\psi_n^{\text{mol}}\rangle,  
\end{equation}
with $\lambda \neq 0$. In other words, the QP orbitals are obtained by diagonalizing the matrix $A$ at its peak energies,
and picking the eigenvector corresponding to the largest eigenvalue (in practice we find the largest eigenvalue is about $10^3$ times
larger than the second largest). We stress that this method of obtaining the QP wave functions is equivalent to solving the QP equation (\ref{eq.QPeq}), but is more convenient for systems with open boundaries.

\begin{figure}[h!]
 \includegraphics[width=0.9\linewidth]{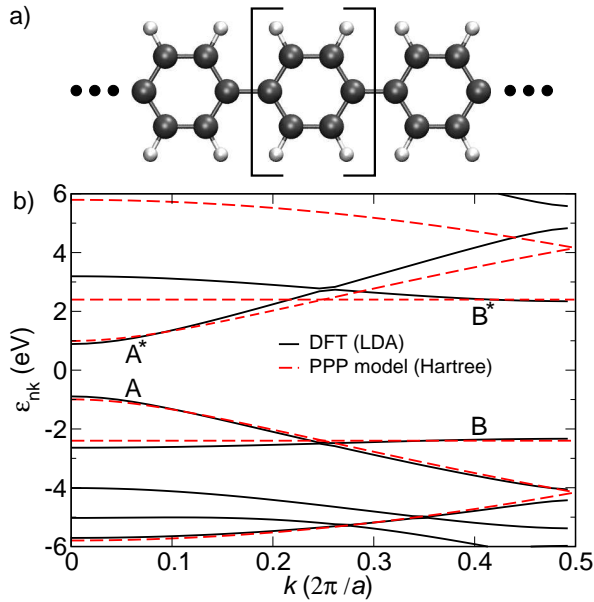}
 \caption{\label{fig1}(Color online) (a) The polyparaphenylene molecular wire. (b) Band structure of the infinite polyparaphenylene molecular wire calculated with DFT(LDA) and the PPP model Hamiltonian with interactions treated at the Hartree level, respectively. The two highest valence bands (A and B) and two lowest conduction bands (A$^*$ and B$^*$) are indicated.}
\end{figure}

\begin{figure*}
  \includegraphics[width=0.95\linewidth]{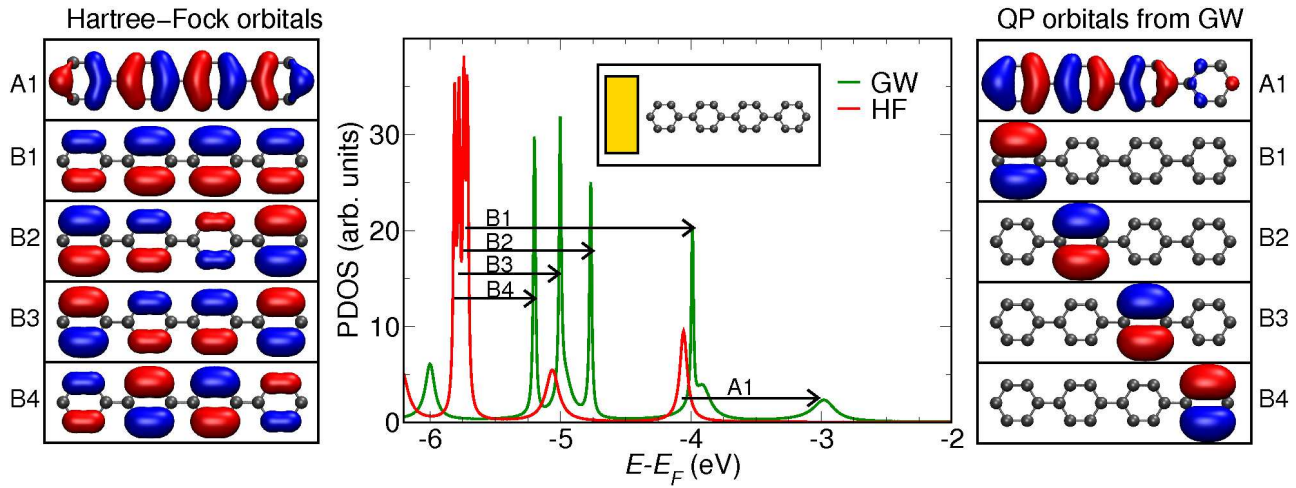}
  \caption{\label{fig2}(Color online) Projected density of states of a 4-unit paraphenylene molecular wire coupled to a metal surface (middle panel). The red and green curves show results obtained at the Hartree-Fock and GW level, respectively. For five energy levels the shift due to correlations (mainly the image charge effect) is indicated by arrows, and the corresponding orbitals are plotted in the left (Hartree-Fock) and right (GW) panels. Orbital A1 is the HOMO and belongs to band A in Fig. \ref{fig1} while the orbitals B1-B4 belong to the narrow band B. The orbitals were constructed by superposing $p_z$ orbitals with weights given by the discrete wave functions of the model.}
\end{figure*}

\section{Results}  \label{sec.results}

In Fig. \ref{fig1} we show the band structure of the infinite poly-paraphenylene wire obtained from a DFT-LDA calculation (red) and the PPP model with interactions described at the Hartree level (blue). We have verified that the LDA xc-potential merely provides a constant shift of all the bands, and thus the two levels of approximation are directly comparable. We conclude that the PPP model yields a reliable description of the $\pi$ bands of poly-paraphenylene. The bands denoted A and A$^*$ are mainly composed of the HOMO and LUMO orbitals of the benzene units, while the narrow B and B$^*$ bands are formed by the HOMO-1 and LUMO+1 benzene orbitals, respectively.

In Fig. \ref{fig2} we show the projected density of states of the 4-unit paraphenylene molecule coupled to the metallic chain 
\begin{equation}
\text{PDOS}(\varepsilon) = \sum_{i\in \text{mol}}A_{ii}(\varepsilon).
\end{equation}
For five molecular levels we indicate the shift due to correlations (mainly the image charge effect) by horizontal arrows. The single-particle orbitals obtained from HF are shown to the left while the QP orbitals from GW are shown to the right. The weight of all the depicted QP orbitals is very close to 1 and they are essentially orthogonal indicating that the single-particle picture applies.

The HF orbitals are completely delocalized over the molecule and are
essentially identical to the orbitals of the freemolecule. This is in
sharp contrast to the orbitals derived from GW which are localized on
different parts of the molecule. The localization of the QP wave
functions occur because of the interaction between the hole on the
molecule and the image charge that it induces in the metal surface.
This is a highly non-local correlation effect and is completely missed
by the HF approximation.

It is clear that the orbitals belonging to the narrower B band become more localized
than the orbitals belonging to the wider A band. This is because it is
energetically cheaper to redistribute the orbitals of a narrow band.
Focusing on the B1-B4 states we observe a clear trend in the
localization: The closer the energy of an orbital is to $E_F$, the
closer to the surface is the orbital localized. Due to the image
charge effect it is always energetically favorable for the hole to
reside closer to the surface. On the other hand, the QP orbitals
should remain (almost) orthogonal, at least when the QP picture
applies as is the case here, and this prevents that all orbitals
contract towards the surface. We recall from Eq. (\ref{eq.QPen1}) that
occupied QP orbitals, $\psi_{\mu}^-$, lying closer
to $E_F$ correspond to many-body excitations, $\Psi_\mu^{N-1}$, with
lower energy. The observed trend in the localization then follows from
the variational principle applied to the may-body states $\Psi_\mu^{N-1}$. 

The unoccupied orbitals are affected by the metal surface in a similar
way with orbitals lying closer to $E_F$ becoming localized more
towards the surface and experiencing a larger energy shift toward $E_F$. The fact that the sign of the image charge shift of the energy of empty and occupied orbitals is different shows that the effect cannot be mimicked by a local $-1/z$ potential. Such a potential would shift all orbitals in the same direction (downwards). To mimick the image charge potential would instead require a non-local potential of the form 
\begin{equation}
\hat V_{img} \sim 1/z \hat P_{\text{occ}}-1/z \hat P_{\text{empty}}
\end{equation}
where $\hat P_{\text{occ}}$ and $\hat P_{\text{empty}}$ project onto the subspace of occupied and unoccupied molecular orbitals, respectively. From this property of the image charge potential, it is clear that the effects presented in Fig. \ref{fig2} cannot be captured by a local potential -- even the exact xc-potential of DFT.

From Fig. \ref{fig2} we see that not only the QP peak positions but also the width of the resonances is affected by the image charge effect. This is particularly pronouned for the A1 orbital which becomes significantly broadened due to the increased weight of the orbital at the carbon atom connected to the metallic chain. For the B orbitals, which have very little weight on the contacting carbon atom, the small increase in the GW peak width relative to HF comes from the (small) imaginary part of the GW self-energy.   
 
\begin{figure}
\includegraphics[width=0.95\linewidth]{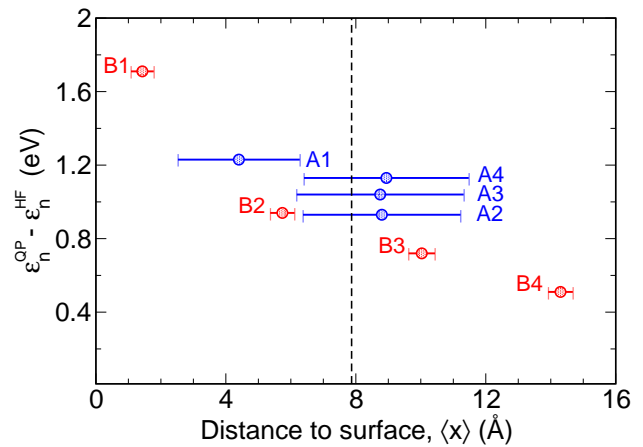}
  \caption{\label{fig3}(Color online) The size of the image charge effect is plotted for the orbitals A1-A4 and B1-B4 against the center of the QP orbital along the axis of the molecule, $\langle x\rangle$. The degree of localization quantified as the second moment $\langle (x-\langle x\rangle)^2\rangle$, is indicated by a horizontal line for each orbital.}
\end{figure}
 
The QP energies include correlations in addition to the exchange
effects described at the Hartree-Fock level. The correlation energy
contains contributions from the Coulomb interactions internally on the
molecule (internal screening) as well as the interactions between metal and molecule (image charge screening). From
calculations for the molecule in the gas-phase we have verified that the contribution from internal screening is almost the same (between 0.4 and 0.6 eV) for the different molecular
orbitals. Hence, apart from this constant, the difference between the QP energy
and the HF energy represents the shift in the energy level due to the image charge effect.

In Fig. \ref{fig3} we plot the image charge shift for the A and B
orbitals against the center of the QP orbital along the axis of the
molecule (the $x$-axis). The center is defined as the first moment, $\langle \psi_n^{\text{mol}}| \hat x| \psi_n^{\text{mol}} \rangle$.  The vertical dashed line indicates the center of the molecule which coincide with the center of the HF orbitals. For each orbital, the degree of
localization, quantified as the second moment $\langle \psi_n^{\text{mol}}|(x-\langle
x\rangle)^2|\psi_n^{\text{mol}}\rangle$, is indicated by a horizontal line. The vertical dashed line indictaes the center of the molecule which coincide with the center 
As expected there is a clear correlation between the size of the image charge shift and the orbital center, in particular for the highly localized B orbitals. The orbitals A2-A4 are all pushed slightly away from the surface.   

\begin{figure}
  \includegraphics[width=0.95\linewidth]{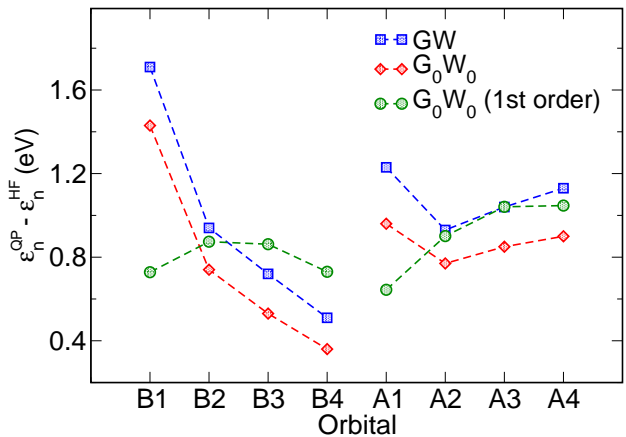}
  \caption{\label{fig4}(Color online) The image charge shift for the A and B orbitals calculated using three different methods for calculating the GW self-energy and solving the QP equation (see text).}
\end{figure}

Fig. \ref{fig4} compares the image charge shifts obtained using
three different strategies for calculating the GW self-energy and solving the QP equation: (i) full solution of the QP equation with a self-consistent
GW self-energy (ii) full solution of the QP equation with a one-shot G$_0$W$_0$ self-energy with $G_0$ from Hartree-Fock, and (iii)
first-order perturbation theory applied to the G$_0$W$_0$ self-energy, i.e. $\varepsilon_n^{\text{QP}}=\varepsilon_n^{\text{HF}}+\langle \psi_n^{\text{HF}}|\Sigma_{G_0W_0}(\varepsilon_n^{\text{HF}})|\psi_n^{\text{HF}}\rangle$. As expected, the first-order approximation (iii) does not perform well in cases where the QP orbitals deviate significantly from the HF orbitals, i.e. for B1-B4 and A1. In particular, for the B1-B4 orbitals the first-order approach predicts similar image charge shifts whereas the shifts obtained with methods (i) and (ii) vary due to the variation in the distance of the QP orbitals to the surface.
As a general trend, the
self-consistent treatment of the GW self-energy leads to larger image
charge shifts (smaller HOMO-LUMO gaps) in agreement with findings for isolated molecules\cite{kaasbjerg}.

At first sight it might seem surprising that the image charge effect, which is essentially electrostatic in nature, is not captured by meanfield theories such as HF. The reason is that the effective potential defining the mean field Hamiltonian does not "know" how an additional particle on the molecule becomes screened by the metal. Under some limiting conditions, however, the
effect of image charge interaction can be simulated using mean field
methods to compute the total energy with an extra particle explicitly
present on the molecule\cite{kaasbjerg,stadler}.

\section{Conclusion} \label{sec.conclusion}

We have discussed the general mathematical and physical meaning of
quasiparticle (QP) wave functions in inhomogeneous systems. For the specific case of a
molecule adsorbed on a metal surface, we found that the QP states can differ
qualitatively from the orbitals obtained within standard
independent-particle approximation as examplified by the Hartree-Fock approximation. Using the GW method, it was shown that image charge
interactions pull the QP frontier molecular orbitals towards the surface. In contrast, the Hartree-Fock single-particle orbitals remain delocalized and identical to those of the isolated molecule.  These results are of importance for the modelling of energy level alignment and
electron transport across metal-molecule interfaces, and should be
observable by low-temperature scanning probe experiments on molecules
on insulating substrates.
 
\section{Acknowledgement}
KST acknowledge support from the Danish Research Council's
Sapere Aude program.
 
\newpage
 
 
\bibliographystyle{unsrt}

\begin{thebibliography}{1}
\bibitem{continuous} For completeness we briefly comment on the
  situation encountered for infinite systems. In this case Eq.
  (\ref{eq.QPeq}) does not admit any exact solutions. Precisely, for
  $\varepsilon$ belonging to the continuous spectrum, the eigenvalues,
  $E_i(\varepsilon)$, of the operator $[H_0 +
  \Sigma_{xc}(\varepsilon)]$ are complex with a finite imaginary part. One can,
  however, define approximate solutions from the condition that
  $\text{Re}E_i(\varepsilon_i)=\varepsilon_i$\cite{kohn-sham66}. These
  solutions represent a group of excitations and can be
  viewed as an average over states of the form defined in Eqs.
  (\ref{eq.QP1},\ref{eq.QP2}) within an energy window around
  $\varepsilon_i$ of width $\text{Im}E_i(\varepsilon_i)$. Solutions
  with small $\text{Im}E_i(\varepsilon_i)$ represent well defined
  excitations and will closely resemble the states in Eqs.
  (\ref{eq.QP1},\ref{eq.QP2}).
\end{thebibliography}

\end{document}